\documentclass[sigconf, nonacm]{acmart}
\AtBeginDocument{%
  \providecommand\BibTeX{{%
    \normalfont B\kern-0.5em{\scshape i\kern-0.25em b}\kern-0.8em\TeX}}}

\usepackage{csquotes}
\usepackage{hyperref}
\usepackage{balance}

\begin{document}





\title{Designing Social Robots that Engage Older Adults in Exercise: \\A Case Study}

\author{Victor Nikhil Antony}
\affiliation{
  \institution{The Johns Hopkins University}
  \city{Baltimore, MD}
  \country{USA}}
\email{vantony1@jhu.edu}

\author{Chien-Ming Huang}
\affiliation{
  \institution{The Johns Hopkins University}
  \city{Baltimore, MD}
  \country{USA}}
\email{chienming.huang@jhu.edu}

\renewcommand{\shortauthors}{Antony and Huang}

\begin{abstract}
We present and evaluate a prototype social robot to encourage daily exercise among older adults in a home setting. Our prototype system, designed to lead users through exercise sessions with motivational feedback, was assessed through a case study with a 78-year-old participant for one week. Our case study highlighted preferences for greater user control over exercise choices and questioned the necessity of precise motion tracking. Feedback also indicated a desire for more varied exercises and suggested improvements in user engagement techniques. The insights suggest that further research is needed to enhance system adaptability and effectiveness to better promote daily exercise. Future efforts will aim to refine the prototype based on participant feedback and extend the evaluation to broader in-home deployments.
\end{abstract}




\begin{teaserfigure}
  \includegraphics[width=\textwidth]{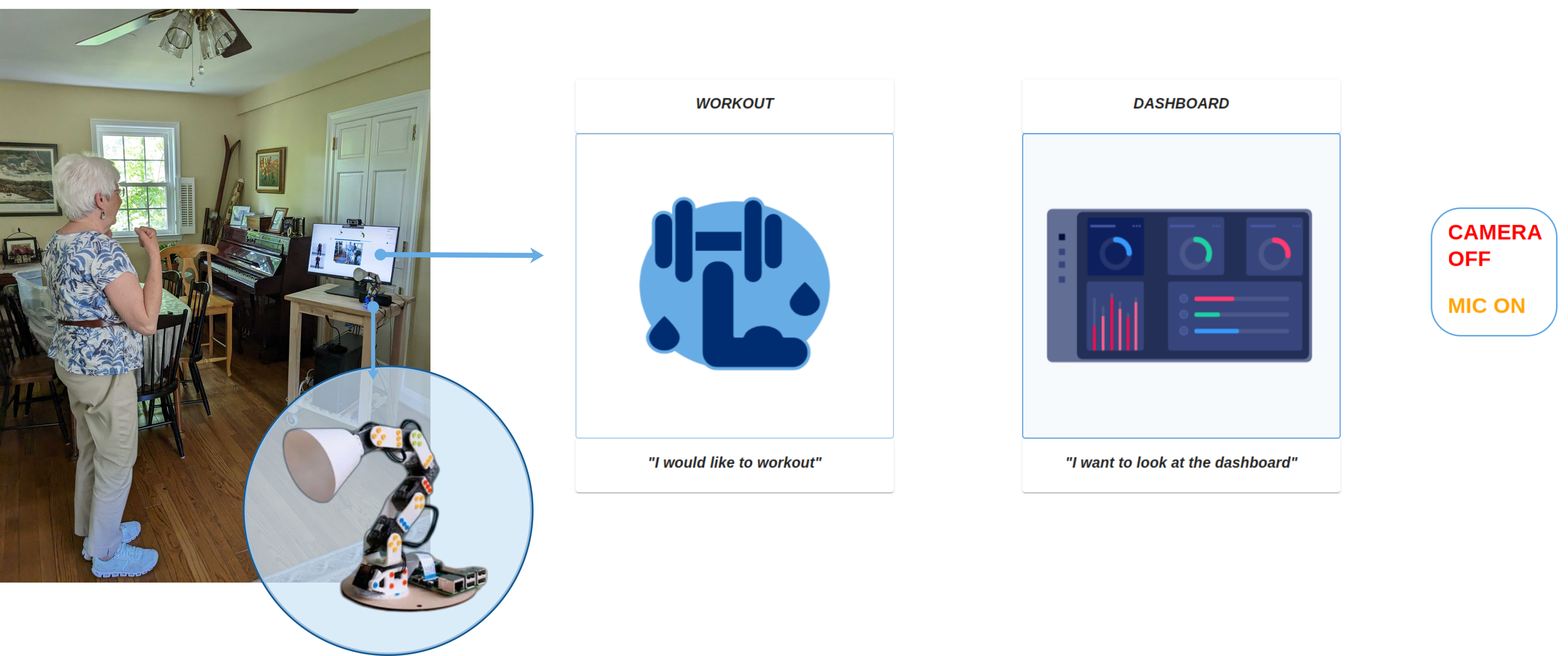}
  \caption{Our prototype features a large touchscreen display along with a Poppy Ergo Jr robot as the embodied social agent}
  \Description{Enjoying the baseball game from the third-base
  seats. Ichiro Suzuki preparing to bat.}
  \label{fig:teaser}
\end{teaserfigure}


\maketitle


\section{Introduction}

As people grow older, regular exercise tends to simultaneously become a tedious yet critical activity linked to a plethora of positive physical and mental health outcomes such as disease prevention and enhanced self-esteem \cite{baum2003effectiveness,dawe1995low,sparling2015recommendations, callow2020mental, toros2023impact}. However, most older adults do not get the recommended levels of regular exercise \cite{us2008physical, us1996physical}. 

Social robots have been proposed to be used as platforms to motivate older adults to engage in physical activity. Several works in this field have explored the design and implementation of such robots and evaluated their effectiveness \cite{antony2023co, lotfi2018socially, fasola2013socially}. However, most of these efforts have been restricted to clinical and laboratory settings and the in-the-wild explorations have been largely restricted to nursing home and short-term settings. There is a need to better understand how to promote exercise in older adults' homes to support independent living and aging in place---a critical factor for a significant portion of older individuals. 

Towards evaluating the impact of and understanding older adults needs from exercise promoting social robots in long-term, in-home scenarios, we developed a prototype social robot platform designed to engage older adults in daily exercise sessions (see Fig. \ref{fig:teaser}). We gained key insights on the viability of our prototype through an in-home case study of our system. We plan on updating our prototype based on the lessons learned from our case study and to use the updated system in our larger efforts in understanding the design of social robots to promote physical activity.


\begin{figure*}[t]
\centering
\includegraphics[width=\textwidth]{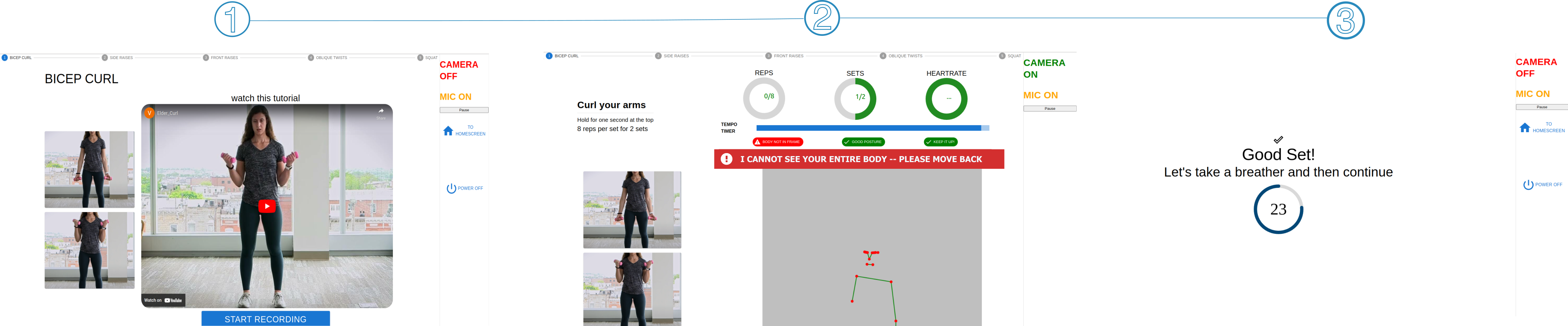}
\caption{Our prototype exercise session begins with (1) showing a demo video of the exercise, and then (2) actively tracks the participant as they complete the set and (3) provides a break between each set.}
\label{fig:workflow}
\end{figure*}

\section{System Overview}

Our prototype system is designed to engage users in daily exercise sessions centered around the social robot's interactions. Our design was informed from insights from consultations with colleagues with expertise in nursing and personal training for older adults. To build our prototype, we also extracted a set of design guidelines from our prior work on co-designing robots to promote physical activity with older adults \cite{antony2023co}. 

For this prototype we focused on supporting goal setting and progress tracking, providing affective and motivational feedback (i.e. encouragement) through engaging dialogue and providing informative feedback (i.e. active tracking, exercise demonstration, benefits of exercises). We chose this subset of features as our prototype fits most closely to the \textit{trainer} role from our prior co-design study \cite{antony2023co}.

Our platform is composed of a Poppy Ergo Jr robot\footnote{Poppy Ergo Jr: \url{https://www.poppy-project.org/en/}}, a large touchscreen display and a computer to run the program (see Fig. \ref{fig:teaser}). Users can interact with the system through speech input which is invoked by saying ``Hey Poppy''; alternatively, users can also use the touchscreen display to interact with the system. Our platform can be linked with a Fitbit Versa 2 smartwatch to allow the system to have access to the participant's real-time bio-stats (e.g. heart-rate). The platform consists of two primary functions: the exercise session and the progress dashboard. 


\subsection{Exercise Session}

At the start of each exercise session, the robot reminds the user to put on their Fitbit, wear comfortable clothing and only exert themselves as they are able. For the purposes of this pilot deployment, the exercise session itself consists only five different exercises (i.e. Bicep Curls, Side Raises, Front Raises, Oblique Twists, and Chair Squats); these exercises were selected in consultation with experts in personal training and nursing. Prior to each exercise, a demonstration video of the exercise is played. After the demo video, the robot leads the user through three sets of the exercise with each set consisting of ten reps. 

During the exercise session, the robot provides motivational feedback (e.g. ``keep going, just three more reps!'', ``you got this! almost there'') during and at the end of each set; the robot includes the user's name during the feedback to build rapport. Between each set, there is a 60-second break. Our robot provides informative feedback during the exercise session by counting the reps out loud for the participant and alerts them if they are doing the exercise incorrectly (see Fig. \ref{fig:workflow}.2) Moreover, the robot is able to provide verbal and visual warnings (e.g. ``slow down a bit, take it easy'') if the participant's heart-rate (measured by a Fitbit smartwatch) is too high


The system performs pose detection using Mediapipe on the video stream and is able to classify the user's pose to count the reps of each exercise. For the pose classification, we trained a K-Nearest Neighbours model that takes in pose landmarks as inputs and outputs the exercise class. 

\subsection{Progress Dashboard}

At the conclusion of each exercise session, users are directed to the dashboard that visually represents their weekly progress. The robot articulates a verbal summary of the user's achievements, designed to prompt reflection and goal setting. Post-session, users can navigate back to the home screen to either conclude their interaction or explore other functionalities of the system.

This dashboard aims to serve as a record of the user's commitment to their exercise journey, reinforcing the behavioural loop of exercise and reward through a visual and interactive medium.

\subsection{Implementation Details}

Our system is built using a ROS2 backend and a ReactJS frontend. We developed Docker containers to simplify the process of replicating our system for future deployments.

\begin{figure}[b]
\centering
\includegraphics[width=\columnwidth]{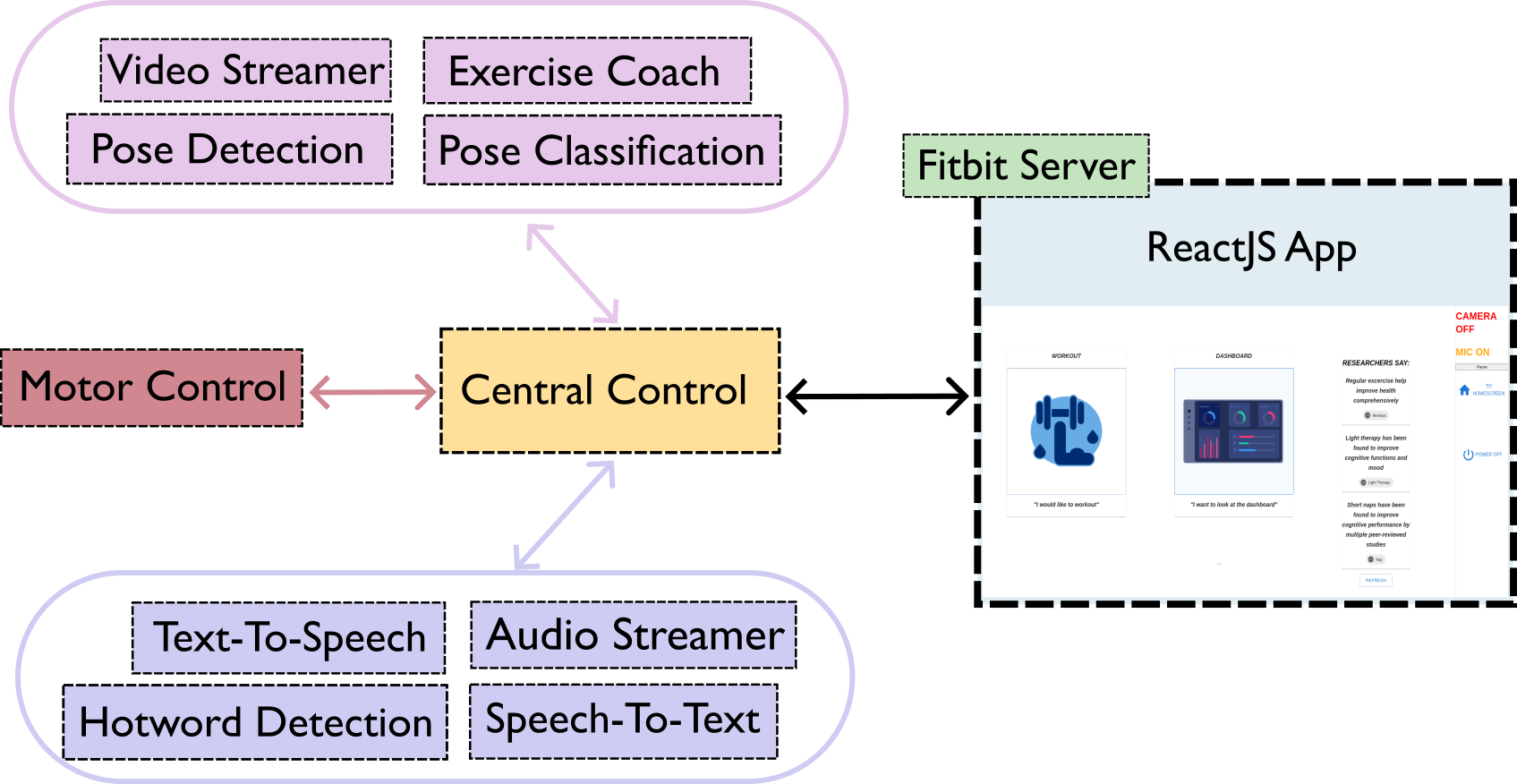}
\caption{Illustration of Our Software Architecture}
\label{fig:soft-arch}
\end{figure}

The ROS2 backend runs modules such as robot motor control, hotword detection, speech recognition, text-to-speech, and pose classification. The ReactJS frontend allows us to render and control the user interface. A central control ROS2 node synchronizes the ROS2 functional modules with the frontend to orchestrate a smooth user experience. See Figure \ref{fig:soft-arch} for an illustration of the software architecture. 

\section{In-Home Case Study}

To evaluate our system, we conducted an in-home case study by deploying the prototype system to the home of an older adult for a duration of seven days.  Our participant was a 78-year-old Caucasian woman living alone at her suburban house with no notable mobility constraints or health conditions. After getting informed consent, we setup the system in the participant's living room and asked the participant to complete the exercise session with the system every day. We conducted a semi-structured interview at the end of the deployment to understand the participant's experience with the system (see Appendix A). 

\subsection{Findings}

Our analysis of the semi-structured interview yielded key insights into the participant's experience and perception of our prototype as well as current limitations of our approach. Our participant engaged with the system for daily exercise all seven days of the deployment. While this adherence was largely driven by study protocol according to the participant, this consistent usage indicates the system's   feasibility with further improvements. Overall, the participant underscored the importance of enhancing user autonomy in choosing exercise protocols and diversifying exercise options to avoid monotony, while questioning the necessity of precise motion detection.

The participant expressed a need for increased autonomy over the exercise protocol in form of options to omit certain exercises, reorder the sequence of activities, and adjust the frequency of exercise sessions to match their preference and lifestyle. 
\begin{displayquote}
\textit{``I don’t want to work out an hour a day, 20 mins a day would be fine but I would prefer it being an hour, 3 days a week.''}
\end{displayquote}

For older adults, the significance of physical activity is derived predominantly from the consistency of exercise rather than the exactness of its execution \cite{amireault2019physical}. Thus, the need for robots accommodate personal preferences and variability cannot be understated in the exercise domain. 

While the participant was initially motivated by the system, a lack of diversity in the exercise offerings led to a perception of monotony. Moreover, the participant did not like the robot saying her name often and instead suggested integrating music indicating that enhancing engagement through multi-modal stimulation may be more effective than singular auditory feedback for sustained user interest. This highlights the need for further exploration in understanding how a social robot can optimally engage and motivate older adults in the exercise domain; for example, leveraging memory of prior interactions to personalize what strategies the robot employs during a session may help enhance the user experience and support continued interaction \cite{baxter2014pervasive}. 


The need for social robots to adapt to users' preferences and effectively engage them becomes increasingly important given that social robotic systems will at least initially occupy a space with existing alternatives. For instance, our participant uses exercise walk-through videos that she finds more engaging and better suited to her needs than our prototype. For social robots to play a meaningful role in supporting an active lifestyle, they must be sufficiently more engaging, helpful and convenient than existing alternatives.

\begin{displayquote}
\textit{``There are so many exercise videos out there that are so nice… I personally don’t need that system… [this instructor I watch] is terrific, she has wonderful exercise and they are 20 minutes exactly and you feel like you have a whole body workout.''}
\end{displayquote}

We also found that instances of erroneous exercise recognition not only disrupted the flow of activity and undermined trust and usage of the platform but also had the potential to aggravate existing physical ailments of the participant. 

\begin{displayquote}
\textit{``I have an arm issue, and unless I completed the exercise exactly the way they [the robot] wanted it done, it wouldn’t move forward so I soldered through.''}
\end{displayquote}

Precision in motion detection and subsequent classification is important to ensure user safety and system reliability however still leads us to question the need for such precise active tracking for the domain. For older adults, it is important to adjust exercises to their capabilities and the focus on building habits for an active lifestyle than fixate on the preciseness of the movement. In this case, perhaps active tracking is not necessary and there may exist simpler mechanisms for a robot to support regular exercise sessions for older adults. 

Replacing the precise and active exercise classification that is currently implemented in the system with a coarser motion tracking system could allow us to to support a wider range of physical activity while ensuring that the user is engaging in exercise

The absence of significant commentary and feedback from the participant on the robot signals that our future research needs to work on harnessing the robot's physical embodiment and non-verbal behaviors to enhance user engagement and promote regular interaction with subsequent iterations of our system. Our decision to employ a non-humanoid robot was strategic, aiming to facilitate larger scale deployments and evaluate the influence of social robots on the physical activity of older adults in prolonged engagements in the future. However, the limited degrees of freedom of the Poppy Ergo Jr robot present challenges in crafting expressive non-verbal communication. Despite these obstacles, there remains potential to position the robot as a central element of the system, thereby leveraging the advantages of an embodied agent to augment the intervention's effectiveness. A viable strategy to enhance user engagement could involve the robot acting as an exercise partner and imitating the exercises performed by users, thereby fostering a sense of companionship. Despite its limited degrees of freedom, our robot could accompany users in their exercise sessions by simulating specific movements through mapping its joints to correlate with particular human body parts, such as engaging a designated motor to signal leg movements during a chair squat or activating a different motor for arm exercises. This approach could allow the robot to provide visual cues that correspond to the user's movements, creating a more interactive and supportive exercise environment. Moreover, adjusting the robot's posture and lighting to reflect emotional states in response to the user's exercise consistency could offer a method of visual feedback, extending beyond the exercise sessions themselves. Enhancing the robot's ability to convey more dynamic and engaging movements could significantly strengthen the relationship with older adults, encouraging sustained physical activity.

Lastly, our participant found the Fitbit excessively cumbersome, using it only during exercise sessions. The Fitbit offers opportunities for essential biostat data collection over longer deployments. Identifying ways to make the Fitbit less cumbersome is crucial for post-hoc analysis of the impacts of the robot-driven intervention on participants' daily activity and health statistics over time.

\section{Next Steps}
We plan on improving our system based on the lessons learned through this initial pilot deployment. In the next phase of our design efforts, we will concentrate on boosting user autonomy in choosing exercise protocols, integrating music, enhancing the robot's motivational feedback, and fostering a workout partner dynamic to increase engagement during exercise sessions.

Then, we plan on conducting more in-home deployments to iteratively improve the design of our system. The long-term goal of this project is to build a system to evaluate the effects of a social robot for exercise support in older adults over longitudinal in-home deployments to support independent living and aging in place. Designing social robots iteratively with direct feedback from older adults can help us identify and implement the set of most critical features required for long-term engagement and consistent usage. Grounding our robot platform's functions in design fundamentals derived from real interactions with older adults is paramount towards using social robots as drivers of meaningful in-home exercise interventions and thereby supporting independent aging.

\begin{acks}
This work was supported by the Malone Center for Engineering in Healthcare at the Johns Hopkins University.
\end{acks}

\balance
\bibliographystyle{ACM-Reference-Format}
\bibliography{sample-base}

\appendix

\section{Interview Questions}

\begin{itemize}
\item How easy was it use the system? From turn on to turn off? Were there any challenges or difficulties you faced while using the robot?
\item Did the system provide clear and helpful guidance for each exercise? Were the exercise videos/images easy to understand and follow?
\item Were there any exercises that you found particularly difficult or confusing? If so, which ones and why? Did you feel the number of reps and sets for each exercise appropriate for your fitness level?
\item Was the robot counting your reps and sets during the exercises helpful?
\item Did you feel motivated and engaged during the exercise session due to the robot’s interactions? Were you excited to use the system to exercise?
\item Did you feel motivated to exercise outside the session due to the interaction with the robot?
\item How was your experience using the Fitbit alongside the exercise robot? 
\item Are there any additional features or functions you would like to see added to the exercise robot?
\item How are you feeling about us taking the system away?
\item Would you be interested in using the exercise robot for a longer period? Would you recommend it to your friends? Why or why not?
\item Is there anything else you would like to share about your experience with the exercise robot?
\end{itemize}

\end{document}